# The FEL SASE operation, bunch compression and the beam heater

*G. Dattoli, M. Labat[1], M. Migliorati[2], P. L. Ottaviani[3], S. Pagnutti[4] and E. Sabia[5]*

ENEA, Unità Tecnica Sviluppo di Applicazioni delle Radiazioni, Laboratorio di Modellistica Matematica, Centro Ricerche Frascati,
C.P. 65 - 00044 Frascati, Rome (Italy)

[1] ENEA Fellow

[2] Università "La Sapienza" Via A. Scarpa 14, 00161 Rome, and LNF-INFN – Frascati

[3] Associato Sezione INFN – Bologna

[4] ENEA, Unità Tecnica Metodi per la Sicurezza dei Reattori e del Ciclo Combustibile, Laboratorio Sviluppo Metodi e Supporto Informatico, Centro Ricerche Bologna

[5] ENEA, Unità Tecnica Tecnologie Portici, Centro Ricerche Portici,
Via Vecchio Macello, 80055 Portici (Napoli)

**ABSTRACT**

We discuss the conditions required for an optimal SASE FEL operation when bunch compression techniques are exploited to enhance the bunch peak current. We discuss the case of velocity bunching and magnetic bunch compression. With reference to the latter technique we provide a quantitative estimate of the amount of laser heater power necessary to suppress the micro-bunching instability without creating any problem to the SASE dynamics.

**Key words**: Free Electron Lasers, Landau Damping, Self Amplified Spontaneous Emission, Synchrotron Light, Electron Beam Accelerators, Coherent Synchrotron Radiation, Michrobunching Instability, Laser Heater

## I. INTRODUCTION

In a SASE FEL device the bunch compression technique is a very useful tool to enhance the peak current, to reach the saturation in a reasonable undulator length. The compression can be achieved using different mechanisms, like

a) The velocity bunching [1]

b) The magnetic bunch compressor [2]

In both cases the compression is accompanied by an increase of the energy spread, which, in the first case, is just due to the longitudinal phase space conservation, while in the second one can be due to coherent synchrotron and/or to micro-bunching instability.

The resulting energy spread may be so large to prevent the SASE process itself, therefore an adequate balance between increase of the peak current and induced energy spread is necessary.

Since the inhomogeneous broadening effects, determining the increase of the gain length, are associated with the relative energy spread, a natural compensation may be determined by an acceleration to higher energies, but this is not always sufficient.

In Figure 1 we have reported the layout of a SASE FEL device, in which a beam of electrons is first passed through a magnetic compressor device, successively it is accelerated to higher energies and eventually used to produce SASE FEL radiation through a chain of undulators. The interplay between michrobunching instabilty, FEL and laser heaters has been recently discussed in depth in a number of dedicated workshops [3].

A laser heater [3-6] is generally used to suppress instabilities of micro-bunching type, but it may also induce a large energy spread preventing the FEL SASE operation in the last part of the system.

The heater solution has been successfully exploited in a recent experiment [7], however its performances were theoretically predicted [8] and experimentally tested in FEL Storage Ring experiments [9], in which it was observed that the on set of the FEL provided a suppression of the Microwave Instability.

The laser heater is essentially provided by a FEL type interaction, in which a laser interacts with the electron bunch inside an undulator, inducing an energy spread. The laser beam may be either external or provided by a superimposed FEL oscillator, as pointed out in ref. [10].

The mechanisms, underlying the suppression, are quite general and they are indeed a by product of the Landau damping. They have been indeed studied for different types of instabilities, including the head tail [11] and the Touscheck beam life time [12], even though the latter cannot be considered, strictly speaking, an instability.

In this paper we discuss the appropriate conditions to have the correct balance between bunch compression and induced spread for an optimal FEL SASE operation. We also study the dependence of the saturated FEL SASE power on the laser heater power, by following the point of view developed in ref. [3].

## II. BUNCH COMPRESSION AND FEL SASE DYNAMICS

Either in ballistics velocity and magnetic bunch compression techniques, the process occurs in different steps: beam compression to increase the peak current, energy increase to compensate the effect of the energy spread induced by the compression itself.

In the case of magnetic compression a further source of energy spread could be induced by the wake fields, which provide contributions associated with the micro-bunching instability.

We will not enter into the mechanisms of the velocity bunching, we will not comment on effect induced by the coherent synchrotron radiation effects, but we will consider the problem from the mere point of view of the high gain dynamics.

To this aim, we remind that the inhomogeneous broadening effect, due to the energy spread, induces an increase of the saturation length, roughly given by [13]

$$\frac{L_S(\tilde{\mu}_\varepsilon)}{L_S} \cong 1 + 0.185 \frac{\sqrt{3}}{2} \tilde{\mu}_\varepsilon^2,$$
$$\tilde{\mu}_\varepsilon = 2 \frac{\sigma_\varepsilon}{\rho}$$

(1)

with $\sigma_\varepsilon$ being the e-beam (uncorrelated) relative energy spread and $\rho$ the FEL SASE gain parameter (Pierce parameter).

According to Fig. 2, where we have reported the ratio between inhomogeneous and homogeneous saturation length as function of $\tilde{\mu}_\varepsilon$, we find that $\tilde{\mu}_\varepsilon = 1$ corresponds to an increase of the saturation length of about 16%, which, for an undulator line of *100 m,* means *16 m* more and, in terms of money, this can be quantified as an extra cost of about one million euros.

We will therefore define a reference upper limit for a correct SASE-FEL operation using the condition

$$\tilde{\mu}_\varepsilon \leq 1 \Rightarrow \sigma_\varepsilon \leq \frac{\rho}{2}$$

(2).

The Pierce parameter scales with the e-beam current density $J$ and energy as

$$\rho \propto \frac{1}{\gamma} J^{\frac{1}{3}} \qquad (3)$$

and $J$ is in turn linked to the peak current $I$ and to the e-beam transverse dimensions by

$$J = \frac{I}{\Sigma},$$
$$\Sigma = 2\pi \sigma_x \sigma_y \qquad (4).$$

The transverse beam sections $\sigma_{x,y}$ can be written in terms of the transverse emittances $\varepsilon_{x,y}$ and of the Twiss coefficients $\beta_{x,y}$, as

$$\sigma_{x,y} = \sqrt{\beta_{x,y} \varepsilon_{x,y}} \qquad (5)$$

and assuming, for simplicity, a round beam, namely a beam with identical emittances and identical Twiss parameters, we find

$$\Sigma = 2\pi \beta \varepsilon_{x,y} \qquad (6)$$

A compression of the electron bunch will determine an increase of the peak current, proportional to the compression factor. In the following we will use the inverse of the compressor factor $\delta$ defined as

$$I_c = \frac{I_o}{\delta} \qquad (7)$$

where the subscripts $0, c$ stand for uncompressed and compressed beam, respectively. The beam compression will create an additional energy spread, which will be compensated by increasing the e-beam energy in the acceleration section.

As already said in the introduction, the increase of the energy spread may be due to the longitudinal phase space conservation: in this case the energy spread increases proportionally to the natural energy spread with the proportionality constant equal to the compressor factor $\delta^{-1}$. The energy spread may, moreover, increase due or to coherent synchrotron radiation or other collective effects: in this case the natural energy spread combines quadratically with the induced one. In both cases we can write

$$\sigma_{\varepsilon,c} = C\sigma_{\varepsilon,0} \tag{8}$$

where $C$ can be a combination of $\delta^{-1}$ and $\sqrt{1+\left(\sigma_{\varepsilon,ci}/\sigma_{\varepsilon,0}\right)^2}$, with $\sigma_{\varepsilon,ci}$ the compression induced energy spread contribution. The final relative energy spread after compression and acceleration can therefore be written as

$$\sigma_{\varepsilon,f} = C\sigma_{\varepsilon,0}\frac{\gamma_0}{\gamma_f}, \tag{9}$$

with $\gamma_f$ the final relativistic factor. Taking also into account that, at higher energies, the emittance is reduced by the angular relativistic effect, we obtain for the final current density the following expression

$$J_{c,f} = \frac{\eta}{\delta}\frac{\beta_0}{\beta_f}\frac{\gamma_f}{\gamma_0}J_0 \tag{10}$$

and $\eta \leq 1$ accounts for possible emittance deterioration due to the compression.

The $\rho$ parameter after the compression and the acceleration, as consequence of eq. (3), can be related to its low energy uncompressed counterpart as

$$\rho_{c,f} = \left( \frac{\eta}{\delta} \frac{\beta_0}{\beta_f} \left( \frac{\gamma_0}{\gamma_f} \right)^2 \right)^{\frac{1}{3}} \rho_0 \tag{11}$$

where $\beta_{f,0}$ are the final and initial Twiss coefficients respectively.

The condition for a safe FEL-SASE operation becomes

$$\frac{2\sigma_{\varepsilon,f}}{\rho_{c,f}} = C \left( \frac{\delta}{\eta} \frac{\beta_f}{\beta_0} \frac{\gamma_0}{\gamma_f} \right)^{\frac{1}{3}} \tilde{\mu}_{\varepsilon,0} < 1$$

$$\tilde{\mu}_{\varepsilon,0} = \frac{2\sigma_{\varepsilon,0}}{\rho_0} \tag{12}$$

from which we find

$$C \leq \left( \frac{\eta}{\delta} \frac{\beta_0}{\beta_f} \frac{\gamma_f}{\gamma_0} \right)^{\frac{1}{3}} \frac{1}{\tilde{\mu}_{\varepsilon,0}} \tag{13}$$

Assuming typical compression parameters, that is $\eta = 0.5$, $\delta = 0.1$, $\frac{\gamma_f}{\gamma_0} = 5$, $\frac{\beta_0}{\beta_f} = \frac{1}{3}$ and supposing that the compression induced energy spread is much smaller than the natural part, we obtain that

$$\tilde{\mu}_{\varepsilon,0} \leq 0.2 \tag{14}$$

which implies a condition on the initial energy spread equivalent to

$$\sigma_{\varepsilon,0} \leq 0.1\rho_0 \tag{15}$$

### III. MAGNETIC BUNCH COMPRESSION AND LASER HEATER

The heater device has been shown to be crucial, for example, for the LCLS SASE laser FEL operation [7] and should be designed to met the following conditions:

a) the laser power should be sufficiently large to inhibit the growth of the instability itself,

b) the induced energy spread should be small enough not to create problems to the FEL operation.

The compromise between the above points determines the amount of the power of the laser dedicated to the heater.

In Figure 3 we have reported two different conception of a laser heater: an external laser or a FEL oscillator, generated by the laser beam itself.

The solution b) is made possible only if the repetition rate of the electron bunches is such that it meets the cavity round trip requirements. If possible, the second scheme may offer some advantages, like the posibility of avoiding synchronization problems between the external laser and the electron bunches.

The concepts we will draw, regarding the amount of power required for the instability suppression, are however independent on the type of laser heater option. In both cases the interaction is of FEL type, it occurs indeed between a laser beam and an electron bunch co-propagating in an undulator, quasi resonant with the laser frequency.

All the physics concerning the FEL-instability interplay can be understood quite easily and can be expressed as it follows: both effects are characterized by a linear gain which is counteracted by the induced energy spread, providing a kind of gain saturation. The dependence of the gain on the energy spread (induced or not) are almost similar in both cases and it is ruled by the function

$$f(\sigma_\varepsilon) \propto e^{-\alpha \sigma_\varepsilon^2} \qquad (16)$$

where $\sigma_\varepsilon$ is the relative energy spread and $\alpha$ is a coefficient depending on whether we are dealing with the dynamics of the instability or that of the FEL.

The process we have considered is the following: the FEL interaction induces an energy spread which does not allow, in the second part of the device (the bending magnets after the undulator), the growth of the instability which in turn would induce a larger energy spread.

If we denote by $\sigma_{FEL}$ the FEL heater induced energy spread, and with $\sigma_I$ the instability induced energy spread and if the beam relative energy spread at the entrance of the undulator is totally negligible, we make the following assumption

$$\sigma_I = \sigma_I(0) e^{-\frac{1}{2}(R \sigma_{FEL})^2} \qquad (17)$$

where $\sigma_I(0)$ is the energy spread induced in absence of any FEL heater effect and $R$ is a coefficient, we will specify in the following.

The total energy spread after the compressor will be therefore

$$\sigma_\varepsilon = \sqrt{\sigma_I^2 + \sigma_{FEL}^2} \tag{18}.$$

The larger is the input laser power, the larger will be the FEL induced energy spread contribution. This can therefore eliminate the instability energy spread contribution, but it may become too large and comparable with $\sigma_I(0)$. To avoid this effect we need the evaluation of an optimum laser heater power.

The dependence of the FEL induced energy spread on the laser power is given by [5]

$$\sigma_{FEL}(X) \cong \frac{0.433}{N} \exp\left[-0.25(\beta X) + 0.01(\beta X)^2\right] \sqrt{\frac{\beta X}{1-e^{-\beta X}} - 1}$$

$$X = \frac{I_h}{I_S} \leq 10, \quad \beta = 1.0145 \frac{\pi}{2} \tag{19}$$

where $I_h$ is the laser heater power and $I_s$ is the FEL heater saturation power[1], linked to the small signal gain and to e-beam power by

$$I_S g_0 = \frac{P_E}{2N} \tag{20}$$

with $N$ the number of undulator periods where the interaction occurs and $P_E$ the power density of the electron beam in the heater.

We can now use quite a simple argument to give a first estimate of the optimum heater power. We assume that $X \ll 1$ in eq. (19) so that

$$\sigma_{FEL}(X) \cong \frac{0.3}{N}\sqrt{\beta X} \qquad (21)$$

---

[1] The FEL saturation intensity defined here has nothing to do with the SASE FEL saturated power, it is indeed a quantity characteristic of FEL oscillators or FEL low gain amplifiers and denotes the laser power halving the small signal gain.

Assuming that the instability contribution can be neglected since it is suppressed by the FEL induced energy spread, we find that after the acceleration, at the entrance of the SASE FEL undulator the e-beam energy spread is just

$$\sigma_{\varepsilon,f}(X) = \frac{\gamma_0}{\gamma_f}\sigma_{FEL}(X) \qquad (22)$$

Imposing the condition $\frac{2\sigma_{\varepsilon,f}(X)}{\rho_{c,f}} < 1$, we find for the laser heater power the following condition, in terms of the SASE saturated power[2]

$$\begin{aligned} &I_h < \xi \rho P_S, \\ &\xi \cong \frac{N\delta}{1.6 g_0}\left(\frac{\gamma_f}{\gamma_0}\right) \end{aligned} \qquad (23)$$

The value reported in eq. (23) is an upper limit, which should be corrected by keeping into account that the contributions of natural energy spread and instability cannot be neglected. It is reasonable that the factor $\xi$, in eq. (23), should be multiplied by a further term so that we can safely assume

$$I_h \cong a\rho P_S \qquad (24)$$

---

[2] It should be stressed that the optical modes cross section in the heater and in the SASE section are different, we should also take into account a correcting term including the mode area sections. This correction does not play any role when we refer to power and not to intensity.

where $a$ is a number of the order of few unities. It is to be stressed that both $I_h$ and $P_S$ are powers, the identity (24) holds if laser and beam sections in the heater and SASE undulators coincide.

The most significant result of this section is that the power of the heater is related, by a fairly simple equation to the SASE saturated power; such a conclusion will be corroborated in the next section by a more accurate argument.

## IV. CONCLUDING REMARKS

In the following we will report a more accurate computation which will essentially confirm the results obtained by means of the previous naive argument.

The calculation we report is based on a semi-analytical method which has been validated with different numerical procedures.

Before entering the details of the discussion, we note that the suppression of the micro-bunching instability induced energy spread is ruled by the coefficient [10]

$$R \cong k_f R_{56},$$
$$k_f \cong \frac{2\pi}{\lambda_f}$$
(25)

where $k_f$ is the modulation wave number and $R_{56}$ is the longitudinal-transverse Twiss coefficient. If we assume as typical value of $R$ few hundreds and $\sigma_I(0) \cong 0.03$, we find that the total energy spread and the FEL induced part versus the dimensionless intensity $X$ has the behaviour shown in Fig. 4 and it is evident that it exhibits a minimum for $X \cong 0.5$.

Just to give an example helpful to swallow the non dimensional form we note that a FEL operating with a small-signal gain coefficient $g_0 \cong 0.2$, $P_E \cong 10^4$ MW X=0.5 corresponds to an intracavity power $P_L \cong 2.5 \cdot 10^2$ MW.

We expect therefore that the optimum FEL SASE performances should occur in correspondence of this minimum. This is better illustrated in Figs. 5. In Figure 5a) we have reported the SASE laser power versus the undulator length, for different values of $X$. With the chosen parameters the optimum heater laser power is around $X^* \cong 0.5$, with a corresponding saturation length of about 31 m. In Figure 5b) we have fixed such a length and we have evaluated the corresponding SASE power as a function of $X$. The maximum corresponding at $X^*$ is just the value yielding the minimum total energy spread.

The value of the heater power corresponding to the minimum of the total energy spread can be inferred from eqs. (18-21) which yield

$$I_h \cong \frac{7N}{g_0 R^2} \ln(R\sigma_\varepsilon(0)) P_E \qquad (26a)$$

The value given by eq. (26) derives from our assumption that $X \ll 1$ which is not strictly fulfilled. A general conclusion is that the optimum value of the dimensionless laser heater power ($X=f$) is always around a value below 1 (0.6 in this case). For this reason we can write[3]

---

[3] If we use eq. (26) the $\chi$ parameter in eq. (26b) reads $\chi \cong \frac{7}{\sqrt{2}} \frac{\delta}{N g_0 (R\rho)^2} \frac{\gamma_0}{\gamma_f} \ln(R\sigma_\varepsilon(0))$

$$I_h \cong \chi \rho P_s,$$

$$\chi = \frac{f\delta}{2\sqrt{2}\, N g_0 \rho^2} \left(\frac{\gamma_0}{\gamma_f}\right) \tag{26b}$$

This result confirms our previous conclusion that the amount of laser heater power is comparable to the SASE saturated power multiplied by the Pierce parameter.

We have mentioned that the laser heater power should be sufficiently large to suppress the micro-bunching instability, but not the SASE gain. We can establish an other inequality yielding a further feeling on the role of the various parameters entering the game.

Equations (25) and (17) allows to conclude that the condition

$$R_{5,6} k_f \sigma_\varepsilon \cong 1 \tag{27}$$

means a significant reduction of the instability induced energy spread.

In terms of laser heater power we obtain from eq. (27)

$$I_h^* \cong \zeta \rho P_S,$$

$$\zeta = 2.5 \frac{N}{(k_f R_{5,6})^2} \left(\frac{\gamma_i}{\gamma_f}\right) \frac{\delta}{g_0 \delta^2} \tag{28}$$

Imposing the condition that such a value be less than the upper limit predicted by eq. (23) we end up with the following inequality

$$\rho \geq \frac{2}{R_{5,6} k_f} \left(\frac{\gamma_0}{\gamma_f}\right) \tag{29}$$

We have mentioned that the results, we have discussed, have been validated numerically; this is strictly true for the SASE dynamics and for the formulae concerning the heater induced energy spread inside the first undulator. The interplay between heater and micro-bunching instability has been modelled using an analytical model. However, if we compare our results with those obtained with an ab-intio numerical procedure, for example the one developed in ref. [14], we observe that the total energy spread as a function of the laser heater power reported in Fig. 4 has the same behaviour of the total energy spread as a function of that after the laser heater reported in the above mentioned paper.

Let us now point out that our analysis has been limited to the case in which the phase space longitudinal distribution is uncorrelated. Significant consequence may occur if we include a correlation term. We write therefore the relevant distribution as

$$f(z,\varepsilon) = \frac{1}{2\pi\Sigma_\varepsilon}\exp\left[-\frac{1}{2\Sigma_\varepsilon}(\gamma_\varepsilon z^2 + 2\alpha_\varepsilon z\varepsilon + \beta_\varepsilon \varepsilon^2)\right] \tag{30}$$

Where $z$ and $\varepsilon = \dfrac{E-E_0}{E_0}$ denote the longitudinal coordinate and the relative energy respectively. The r. m. s. bunch length and relative energy spread are defined in terms of the Twiss parameters and of the longitudinal emittance $\Sigma_\varepsilon$ as[4]

$$\sigma_\varepsilon = \sqrt{\gamma_\varepsilon \Sigma_\varepsilon},$$
$$\sigma_z = \sqrt{\beta_\varepsilon \Sigma_\varepsilon}, \tag{31}.$$
$$\beta_\varepsilon \gamma_\varepsilon = 1 + \alpha_\varepsilon^2$$

The modification induced by the correlation term in the high gain FEL small signal dynamics can easily be computed using the FEL high gain equation [15], which can be written as

$$\partial_\tau a(z,\tau) = i\pi g_0 \int_0^\tau e^{i\nu\tau' - \frac{(\pi\pi_\varepsilon\tau')^2}{2}} \Phi(z+\Delta\tau,\tau') a(z+\Delta\tau,\tau-\tau') d\tau',$$

$$\Phi(z,\tau) = \exp\left\{-\frac{1}{2}\left[\left(\frac{z}{\sigma_z} + i\pi\mu_{\varepsilon,c}\tau\right)^2 + \mu_{\varepsilon,c}^2\tau^2\right]\right\}, \qquad (32)$$

$$\mu_{\varepsilon,c} = 2N\alpha$$

where $\Delta = N\lambda$ is the slippage length and $\mu_\varepsilon = 2N\sigma_\varepsilon$.

The evolution dynamics is strongly modified by the presence of an energy position correlation term, which provides some extra-contribution which affects either the detuning and the lethargic behaviour.

This is clearly shown in Fig. 6, where we have reported the evolution of an initially Gaussian pulse $f(z) = \frac{1}{\sqrt{2\pi}\sigma_z} e^{-z^2/2\sigma_z^2}$ undergoing a FEL interaction. We have considered the intermediate gain regime only for two cases with and without correlation. The correlated evolution is indeed provided by a lower gain and by a shift of the packet centre of "mass" in the forward direction.

---

[4] The dimensions of the longitudinal emittance $\Sigma_\varepsilon$ are just matter of convention, if we measure $z$ in meters and $\sigma_\varepsilon$ dimensionless, $\Sigma_\varepsilon$ can be expressed in $mm \cdot mrad$ as a consequence $\beta_\varepsilon$ is expressed in $mm$  $\gamma_\varepsilon$ in $mm^{-1}$ and $\alpha_\varepsilon$ is dimensionless.

This problem will be however treated more carefully elsewhere, by means of a three-dimensional extension of eq. (29) in which we will include the full six dimensional phase space and the relevant correlation.

**FIGURE CAPTIONS**

Fig. 1  SASE FEL device with a bunch compressor.

Fig. 2  Ratio between inhomogeneus and homogeneus saturation length vs $\tilde{\mu}_\varepsilon$

   R.F.≡ Radio-frequency cavities

   B.C. ≡ bunch compressor chicane

   U≡ undulator lines

Fig. 3  Laser heater schemes a) external laser (E.L.) b) FEL heater, M≡ mirror

Fig. 4  Total (continuous line) and FEL induced (dot line) energy spread with $\sigma_\varepsilon(0) \cong 3 \cdot 10^{-2}$, $\lambda_{u,heater} \cong 2.15\,cm$, $K = 2$, $N = 50$, $\gamma \cong 200$ (heater laser wavelength $800\,nm$), $\frac{\gamma_f}{\gamma_0} \cong 10$ $\rho \cong 9 \cdot 10^{-4}$.

Fig. 5  a) SASE FEL power vs. the undulator coordinate, for different values of the laser heater power, red $X = 0$, blue $X=0.2$, green $X=0.7$; b) FEL SASE power at $z = 31\,m$ vs. $X$. Same parameters of Fig. 4. The blue line represents the total energy spread of Fig. 4.

Fig. 6  Optical pulse vs. $\frac{z}{\sigma_z}$ at $\tau = 1$ for $g_0 \cong 2, \nu = 1, \frac{\Delta}{\sigma_z} = 1, \frac{\sigma_z}{\sigma_r} = 1, \mu_\varepsilon = 0.1$, red $\alpha_\varepsilon = 4$, blue $\alpha_\varepsilon = 0$.

**Figure(s)**

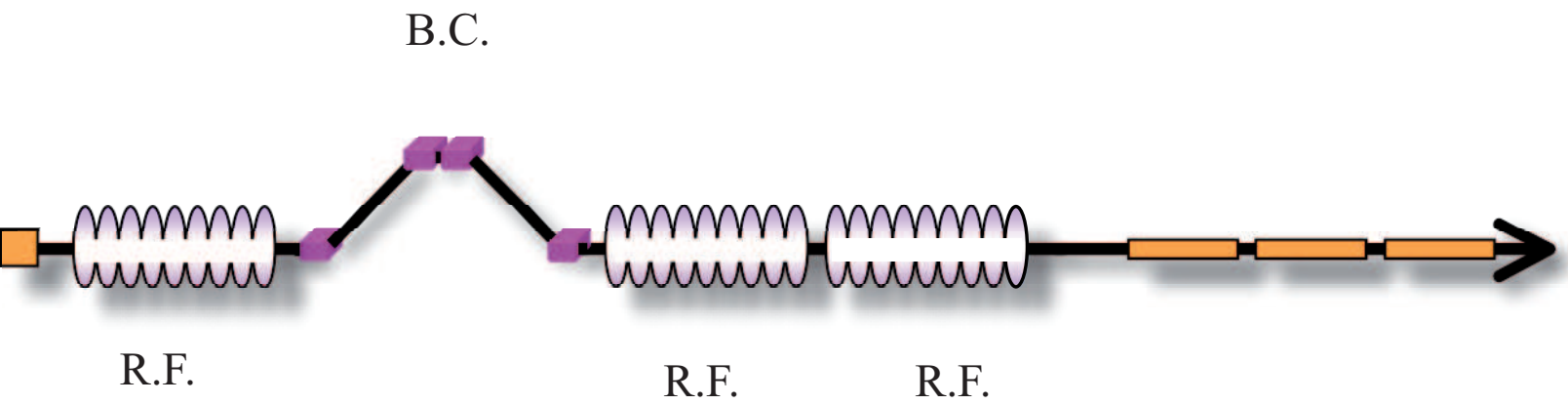

Fig. 1

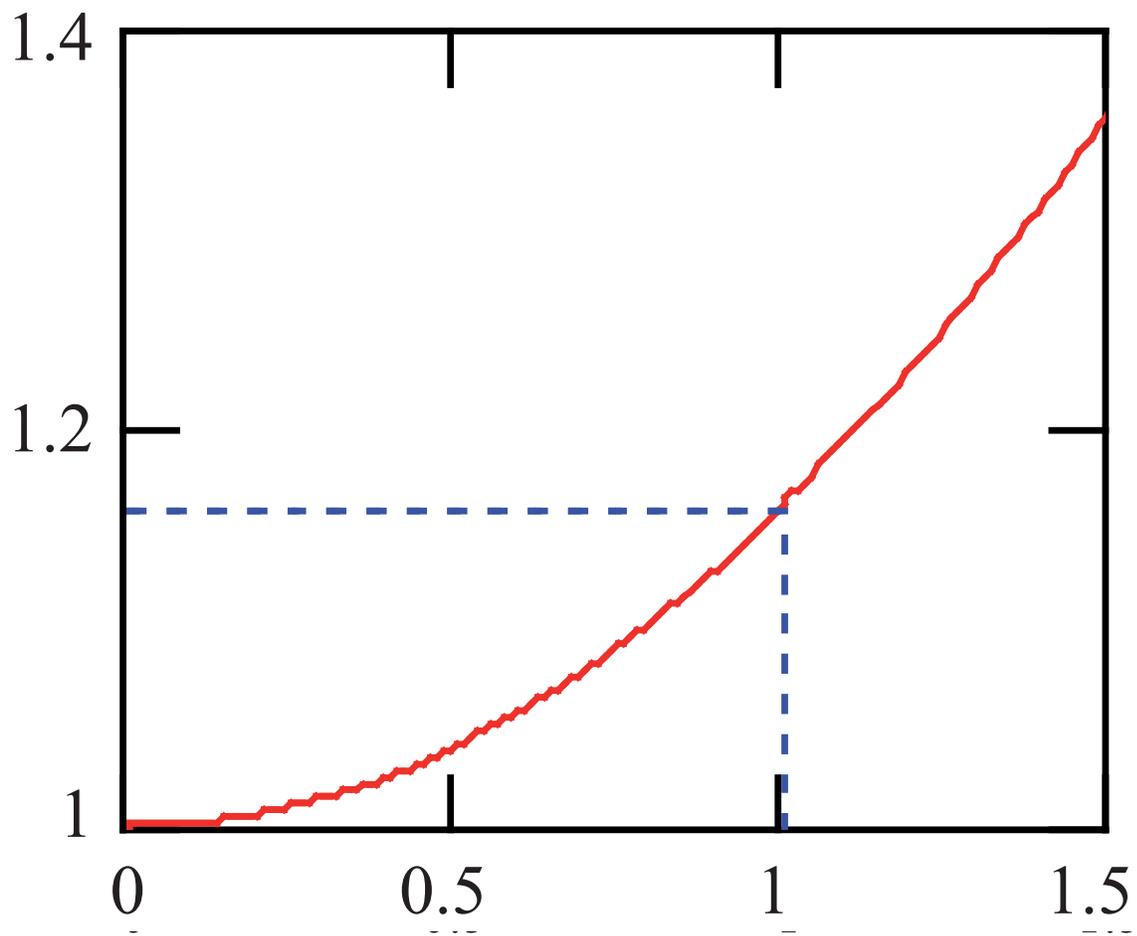

Fig. 2

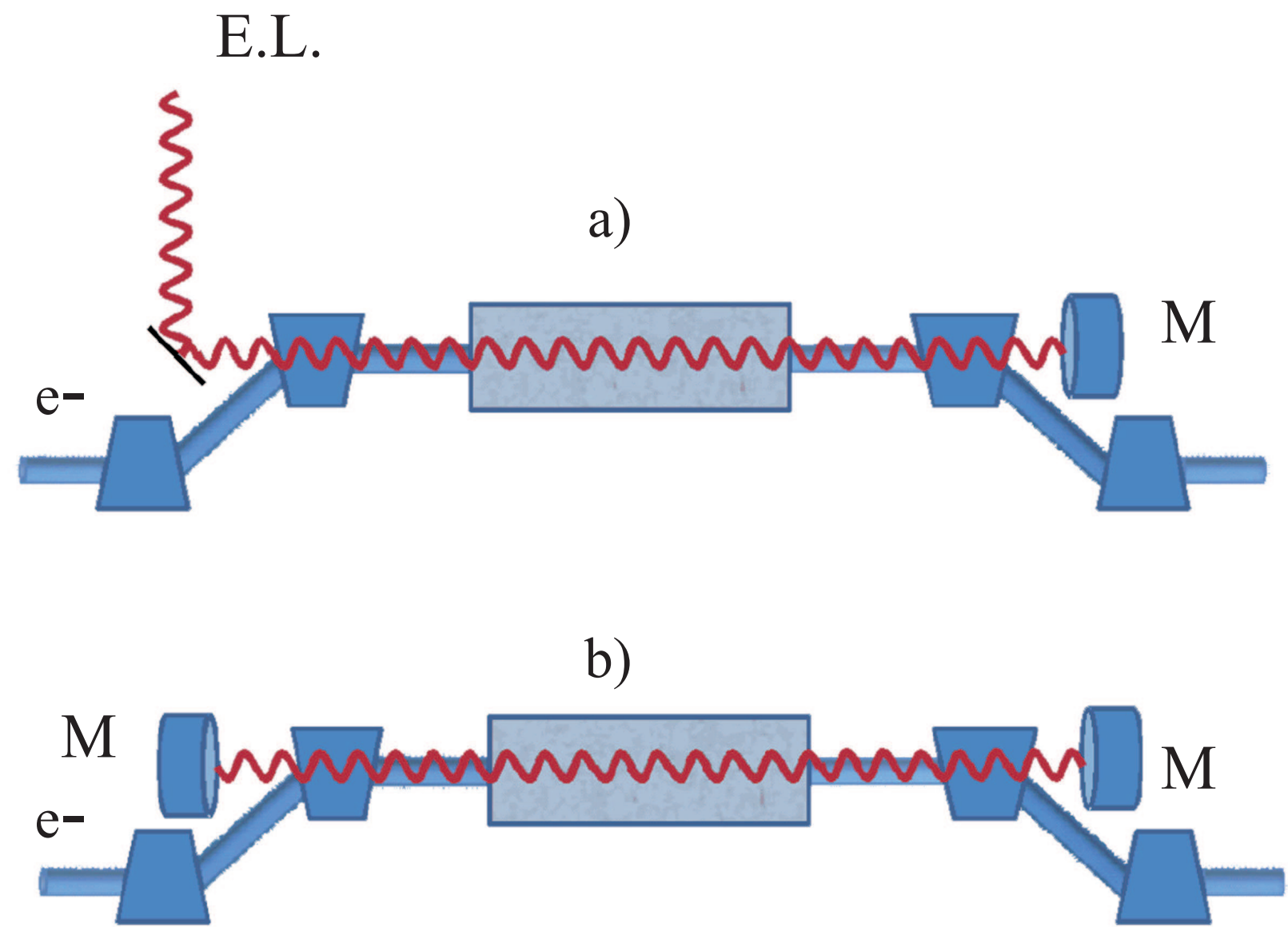

Fig. 3

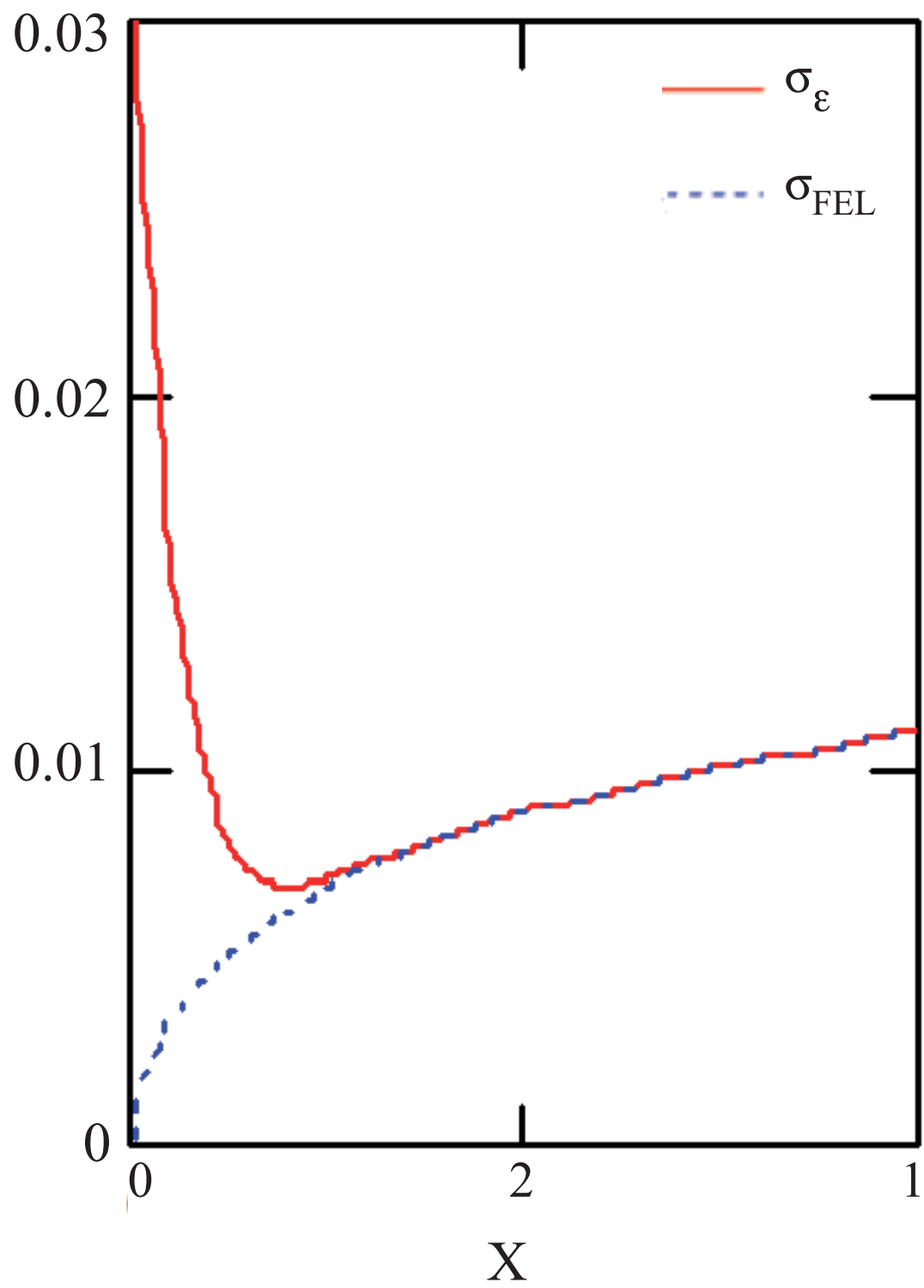

Fig. 4

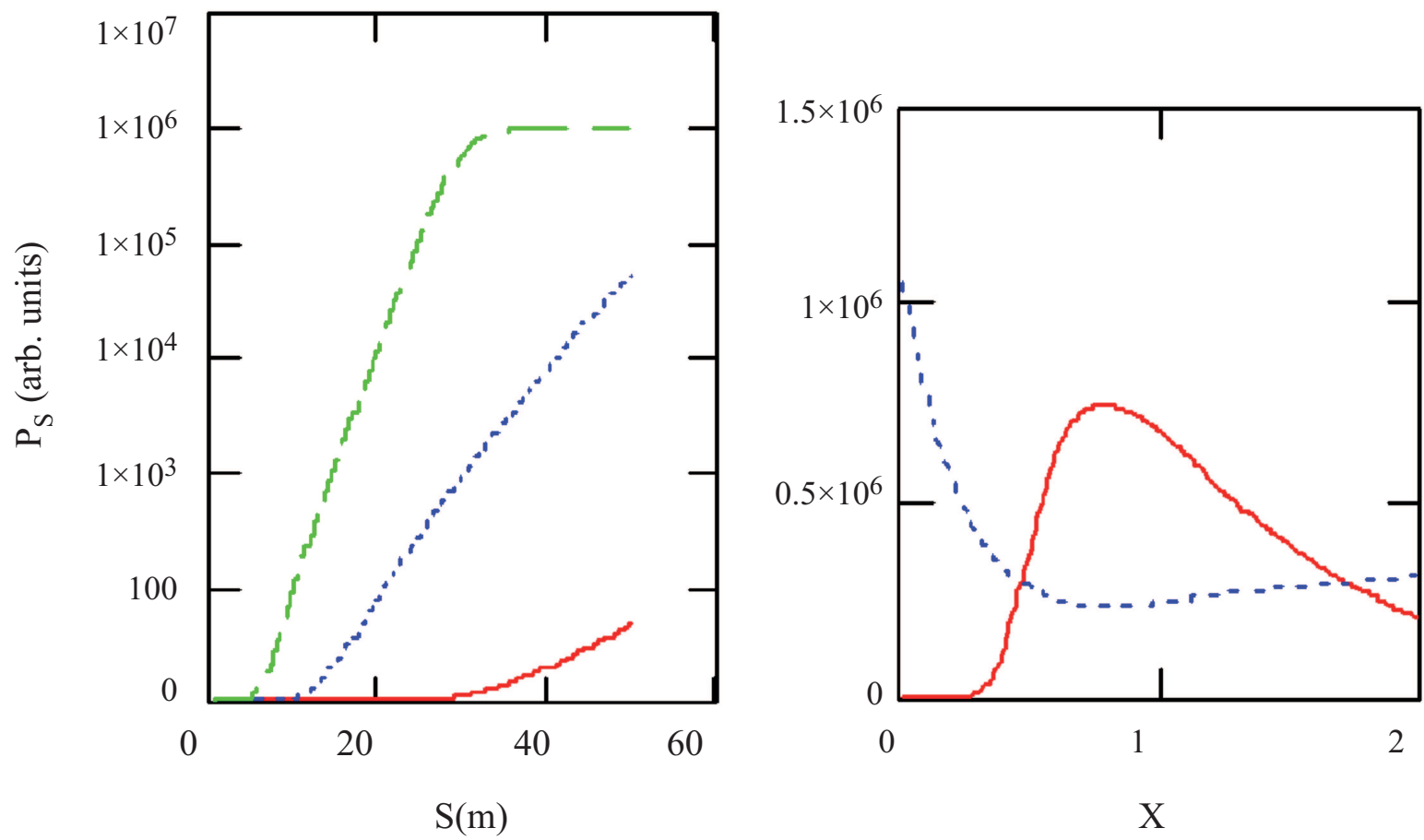

Fig. 5

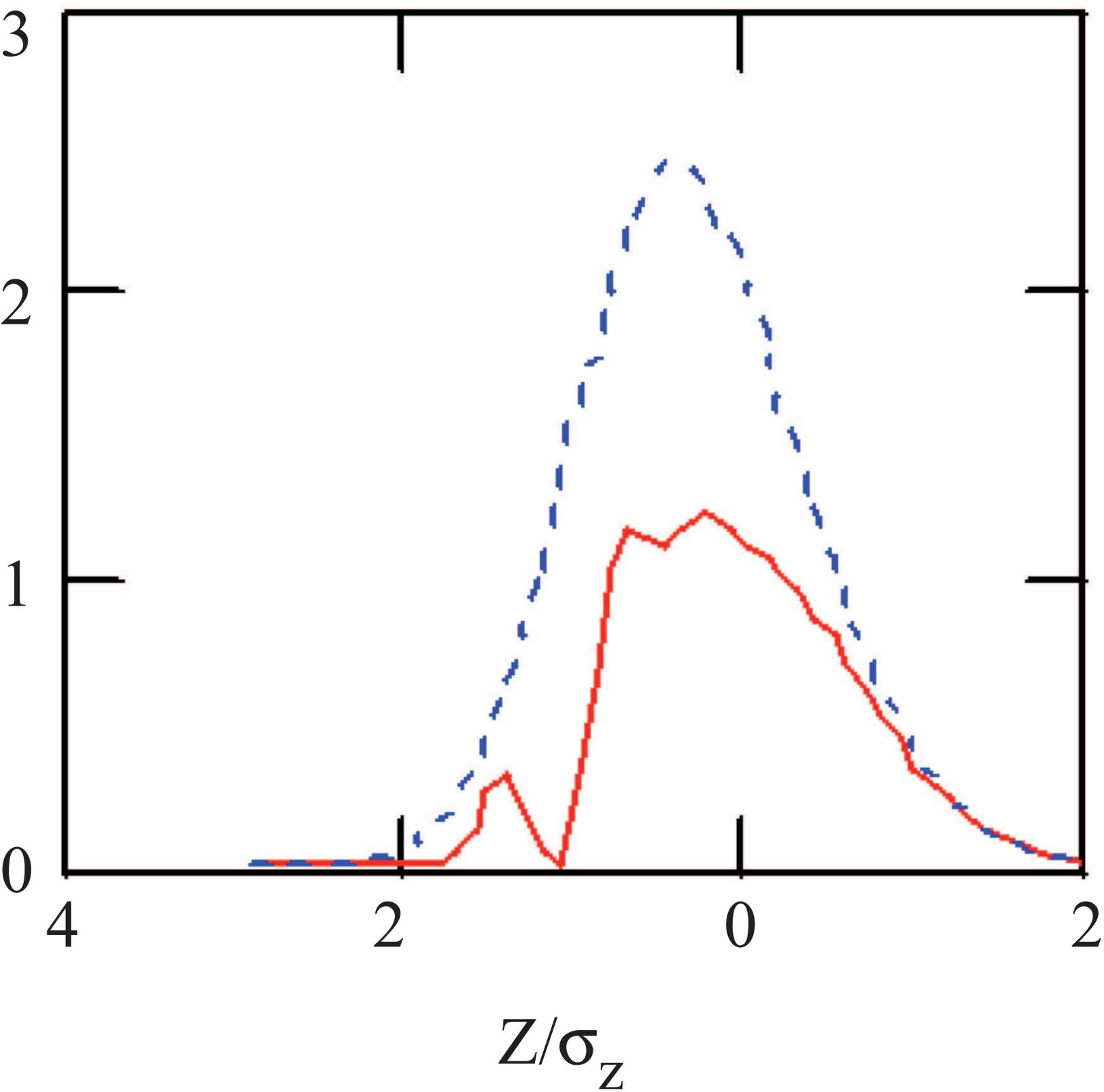

Fig. 6